\documentclass[aps,twocolumn, amsmath,showpacs,amssymb,prl]{revtex4-1}
 
\usepackage{epsfig}
\usepackage{graphicx}
\usepackage{dcolumn}
\usepackage{bm}

\newcommand{\be}{\begin{equation}}
\newcommand{\ee}{\end{equation}}
\newcommand{\ba}{\begin{eqnarray}}
\newcommand{\ea}{\end{eqnarray}}
\newcommand{\nn}{\nonumber}

\newcommand{\psat}{p_{\rm{sat}}}
\newcommand{\ptr}{p_{\rm{T}}}
 
\begin{document}

\title{Fluid dynamics with saturated minijet initial conditions \\in ultrarelativistic heavy-ion collisions}

\author{R.~Paatelainen${}^{a,b}$, K.~J.~Eskola${}^{a,b}$, H.~Niemi${}^{a,b}$, K.~Tuominen${}^{c,b}$ }
\affiliation{$^{a}$Department of Physics, P.O.Box 35, FI-40014 University of
Jyv\"askyl\"a, Finland}
\affiliation{$^b$Helsinki Institute of Physics, P.O.Box 64, FI-00014 University of
Helsinki, Finland}
\affiliation{$^{c}$Department of Physics, P.O.Box 64, FI-00014 University of
Helsinki, Finland}
 
\begin{abstract}
Using next-to-leading order perturbative QCD and a conjecture of saturation to suppress the production of low-energy partons, we calculate the initial energy densities and formation times for the dissipative fluid dynamical evolution of the quark-gluon plasma produced in ultrarelativistic heavy-ion collisions. We identify the framework uncertainties and demonstrate the predictive power of the approach by a good global agreement with the measured centrality dependence of charged particle multiplicities, transverse momentum spectra and elliptic flow simultaneously for the Pb+Pb collisions at the LHC and Au+Au at RHIC. In particular, the shear viscosity in the different phases of QCD matter is constrained in this new framework simultaneously by all these data.
\end{abstract}

 
\maketitle 

The main goal of ultrarelativistic heavy-ion collisions at the Large Hadron Collider (LHC) and the Relativistic Heavy-Ion Collider (RHIC) is to determine the thermodynamic and kinetic properties of strongly interacting matter. The measured hadronic transverse momentum ($\ptr$) spectra at the LHC and RHIC  provide convincing evidence for a formation of a strongly collective system and a nearly thermalized quark-gluon plasma (QGP) \cite{Heinz:2013th}. In particular, the observed systematics of the  Fourier harmonics $v_n=\langle\cos(n\phi)\rangle$ of the azimuth-angle distributions, are remarkably consistent with a low-viscosity QCD matter whose expansion and cooling are describable with dissipative relativistic fluid dynamics \cite{Romatschke:2007mq,Luzum:2008cw,Schenke:2010rr,Gale:2012rq,Song:2010mg,Song:2011qa,Shen:2010uy,Bozek:2009dw,Bozek:2012qs, Niemi:2011ix, Niemi:2012ry}. 

The essential inputs to the fluid dynamics are the initial energy density and flow of the matter created in the collision. However, the final state observables like multiplicities, $\ptr$ spectra and $v_n$, are also strongly affected through the fluid dynamical expansion by the viscosity and the equation of state (EoS). Thus the entire spacetime evolution, including partons in the colliding nuclei, the primary production and thermalization of QCD matter and the subsequent fluid dynamical evolution, becomes highly convoluted.
Description of all these dynamics in a coherent way, leading to quantitative predictions and a meaningful determination of the QCD matter properties from the measurements, provides an ultimate challenge in the field. As discussed in this paper, the determination of e.g. the temperature dependence of the shear viscosity-to-entropy ratio $\eta/s(T)$ calls for a simultaneous theory analysis of all possible bulk (low-$\ptr$) observables at the LHC and RHIC.

Parton saturation is a viable mechanism to control the otherwise unsuppressed production of soft small-$\ptr$ quanta  in hadronic and nuclear collisions \cite{Gribov:1984tu,Mueller:1985wy,McLerran:1993ni,Eskola:1996ce}. In essence saturation means that there exists a semihard scale controlling the particle production in the collision.
In the  perturbative QCD (pQCD) + saturation framework we consider here, the primary particle production in $A$+$A$ collisions is computed in collinear factorization by approaching the saturation at semi-hard scales from the perturbatively controllable high-$\ptr$ side \cite{Eskola:1999fc,Paatelainen:2012at}. Perturbative QCD provides 
an excellent description of hard processes in hadronic and nuclear collisions  at interaction scales $Q\gtrsim 1$~GeV \cite{Eskola:2009uj}.
Moreover, this framework allows for a quantification of the particle production uncertainties, and their propagation through the fluid dynamical evolution in nuclear collisions \cite{Paatelainen:2012at}. In addition to the internal consistency of the pQCD-based approach, it should be noted that perturbative primary gluon production in heavy-ion collisions is complementary to the Color-Glass Condensate  models \cite{Gelis:2010nm}  which build on soft gluon fields. If these different high-energy QCD approaches produce similarly successful heavy-ion phenomenology, the overall uncertainty in determining the QCD matter properties can be dramatically reduced.

The present work has roots in the so-called EKRT saturation model \cite{Eskola:1999fc}, which successfully predicted the multiplicities and $\ptr$ spectra in central $A$+$A$ collisions at RHIC and LHC \cite{Eskola:2001bf,Eskola:2002wx,Eskola:2005ue,Renk:2011gj}, and also the centrality dependence at RHIC \cite{Eskola:2000xq} (cf. Fig.~23(a) in \cite{Abelev:2008ab}).  
Here we use the next-to-leading-order (NLO)-improved pQCD + saturation framework of \cite{Paatelainen:2012at} to calculate the initial QGP energy density profiles and formation times, and combine these with viscous fluid dynamics. We analyse the centrality dependence of charged particle multiplicities,  $\ptr$ spectra and elliptic flow ($v_2$)  at the LHC and RHIC in terms of the few physical key-parameters of the framework. We show that a good simultaneous description of all these observables can indeed be obtained without retuning the framework from one collision system (cms-energy, nuclei, centrality) to another. This results in the robust predictive power of the approach, originating from the pQCD calculation of the QGP initial conditions. Most importantly, this predictive power enables us to study and restrict the ratio  $\eta/s(T)$ in the different QCD-matter phases more consistently in a simultaneous multiobservable analysis of the LHC and RHIC data. 

Let us then discuss the details of our framework \cite{Paatelainen:2012at}.
The rigorously calculable part is the minijet $E_T$ production in an $A$+$A$ 
collision, in a rapidity interval $\Delta y$ and above a $\ptr$ scale $p_0$,
\begin{equation}
\frac{dE_T}{d^2{\bf s}} = T_A({\bf s} + \frac{{\bf b}}{2})T_A({\bf s} - \frac{{\bf b}}{2})\sigma\langle E_T \rangle_{p_0,\Delta y,\beta},
\label{eq: dET}
\end{equation}
where ${\bf s} =(x,y)$ is the transverse location, ${\bf b}$ the impact parameter,  
and $T_A({\bf s})$ the standard nuclear thickness function with the Woods-Saxon nuclear density profile. The first $E_T$-moment of the minijet $E_T$ distribution, $\sigma\langle E_T \rangle_{p_0,\Delta y,\beta}$, \cite{Eskola:1988yh,Paatelainen:2012at} is in NLO
\begin{equation}
\label{eq: sigmaet}
\sigma\langle E_T \rangle_{p_0,\Delta y,\beta} = \sum_{n=2}^{3}\frac{1}{n!}\int [{\rm DPS}]_n\frac{{\rm d}\sigma^{2\rightarrow n}}{[{\rm DPS}]_n} \tilde{S}_n,
\end{equation}
where ${\rm d}\sigma^{2\rightarrow n}$ are the collinearly factorized minijet production cross sections and $[{\rm DPS}]_n$ denote the phase-space differentials for the $2\rightarrow 2$ and $2\rightarrow 3$ cases \cite{Paatelainen:2012at,tuominen:2000}.
We apply the CTEQ6M parton distribution functions (PDFs) \cite{Pumplin:2002vw} with the EPS09s impact-parameter dependent nuclear PDFs \cite{helenius:2012wd}. The measurement functions $\tilde{S}_2$ and $\tilde{S}_3$ define the hard scattering in terms of the minijet transverse momenta  $p_{T,i}$ and the cut-off scale $p_0$, as well as the total minijet $E_T$ produced in $\Delta y$:
\begin{equation}
\tilde{S}_n = \Theta(\sum_{i=1}^n p_{T,i}\geq 2p_0) E_{T,n} \Theta(E_{T,n} \geq \beta p_0),
\end{equation}
where $E_{T,n} = \sum_{i=1}^n \Theta(y_i\in \Delta y)p_{T,i}$ and $\Theta$ is the step function.
These functions, analogous to the jet definitions \cite{Kunszt:1992tn}, are constructed so that $\sigma\langle E_T \rangle_{p_0,\Delta y,\beta}$ is a well-defined, infrared- and collinear-safe, quantity to compute.
The hardness-parameter $\beta$ defines the minimum $E_T$ in the interval $\Delta y$. As discussed in \cite{Paatelainen:2012at}, any $\beta \in [0,1]$ is acceptable for the rigorous NLO computation.

Following the new angle in formulating the minijet saturation \cite{Paatelainen:2012at}, the $E_T$ production is expected to cease when the $3\to 2$ and higher-order partonic processes start to dominate over the conventional $2\to 2$ processes. For a central collision of identical nuclei of radii $R_A$ this leads to a transversally averaged saturation criterion  $E_T(p_0,\sqrt{s_{NN}},\Delta y,\beta) = K_{\rm{sat}} R_A^2 p_0^3 \Delta y$, with an unknown, $\alpha_s$-independent, proportionality constant $K_{\rm{sat}}\sim 1$.
Generalizing to non-zero impact parameters and localizing in the transverse coordinate plane gives
\begin{equation}
\frac{{\rm d}E_T}{{\rm d}^2{\bf s}}(p_0,\sqrt{s_{NN}},\Delta y,{\bf s},{\bf b},\beta) = \frac{K_{\rm{sat}}}{\pi} p_0^3 \Delta y,
\label{eq:localsaturation}
\end{equation}
where the l.h.s. is the $p_0$-dependent NLO pQCD calculation defined in Eq.~\eqref{eq: dET}. 

For given $K_{\rm{sat}}$ and $\beta$, we solve the above equation for $p_0=p_{\rm sat}(\sqrt{s_{NN}}, A, \mathbf{s}, \mathbf{b}; K_{\rm sat}, \beta)$, and obtain the total  ${\rm d}E_T/{\rm d}^2{\bf s}$ in a mid-rapidity unit $\Delta y=1$ at saturation from the r.h.s. as  $K_{\rm{sat}} p_{\rm sat}^3 /\pi$. Once the solution $p_{\rm sat}$ is known, the local energy density is obtained \cite{Eskola:2001bf,Eskola:2005ue} as
\begin{equation}
\varepsilon(\mathbf{s}, \tau_s = 1/p_{\rm sat}) = \frac{dE_T}{d^2{\bf s}\tau_s\Delta y} = \frac{K_{\rm{sat}}}{\pi}p_{\rm sat}^4.
\label{eq:edensity}
\end{equation}
where the local formation time is $\tau_s = 1/p_{\rm sat}$.

\begin{figure}[!h]
\hspace{-0.5cm} 
\epsfysize 5.4cm \epsfbox{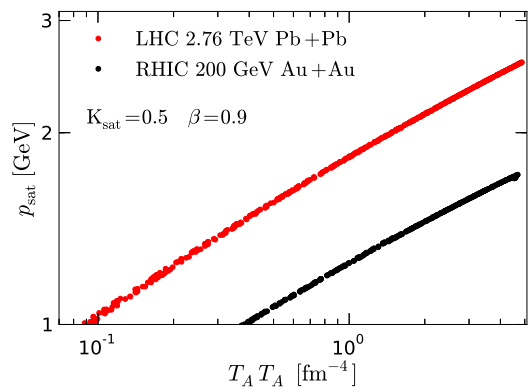} 

\vspace{-0.5cm}
\caption{\protect (Color online) Examples of the saturation momenta obtained 
for the LHC and RHIC $A$+$A$ collisions, as functions of $T_AT_A$. 
See text for details. 
}
\label{fig:fitsaturation}
\end{figure}

Fig.~\ref{fig:fitsaturation} shows examples of $p_{\rm sat}(\sqrt{s_{NN}}, A, \mathbf{s}, \mathbf{b}; K_{\rm sat}, \beta)$ as a function of $T_A T_A$, calculated for fixed values of $K_{\rm sat}, \beta$ and with $b=0$ and three other fixed impact parameters corresponding to the centrality classes 0-5\%, 20-30\% and 40-50\% in $\sqrt{s_{NN}}=2.76$~TeV Pb+Pb collisions at the LHC and 200 GeV Au+Au at RHIC. To a very good approximation, the \textbf{b} and \textbf{s} dependence of $\psat$ comes only through $T_A T_A$. This is due to the weak $\mathbf{s}$ dependence of the nPDFs near the centres of the nuclei \cite{helenius:2012wd}.  The approximate power-law scaling behaviour seen at large $T_AT_A$ can then be understood as expained in \cite{Eskola:2001rx}.

We identify two main uncertainties in mapping the pQCD + saturation
calculation to an initial state for fluid dynamics: (\emph{i}) The energy
density given by Eq.~\eqref{eq:edensity} is at a time $\tau_s = 1/p_{\rm sat}$,
\emph{i.e.} different at each transverse point \textbf{s}, while for fluid dynamics we
need the initial condition at a fixed time $\tau_0$. (\emph{ii}) We cannot
trust the pQCD calculation down to  $p_{\rm sat} \rightarrow 0$, but we 
need to set a minimum scale $p_{\rm sat}^{\rm min} \gg \Lambda_{\rm QCD}$.
Wherever $p_{\rm sat} \ge p_{\rm sat}^{\rm min}$ we can use the pQCD calculation,
but the other regions, \emph{i.e.} low density edges, need to be treated separately.

We fix a minimum saturation scale as $p_{\rm sat}^{\rm min} = 1$~GeV. Correspondingly, 
the maximum formation time in our framework is $\tau_{\rm 0} = 1/p_{\rm sat}^{\rm min}$.
Then, we evaluate the energy densities from $\tau_s({\bf s})$ to $\tau_{\rm 0}$ using either the Bjorken free streaming  
$\varepsilon(\tau_{\rm 0}) = \varepsilon(\tau_s)(\tau_s/\tau_{\rm 0})$ (FS) or the Bjorken hydrodynamic scaling solution $\varepsilon(\tau_{\rm 0}) = \varepsilon(\tau_s)(\tau_s/\tau_{\rm 0})^{(4/3)}$ (BJ). We take these two limits to represent the uncertainty in the early pre-thermalization evolution: In the free streaming case the transverse energy is preserved, while the other limit corresponds to the case where a maximum amount of the transverse energy is reduced
by the longitudinal pressure. 

To obtain the energy density $\varepsilon({\bf s},\tau_0)$ in the transverse region where $p_{\rm sat} < p_{\rm sat}^{\rm min}$, we use an interpolation $\varepsilon = C (T_A T_A)^n$, where the power $n = \frac{1}{2}\left[(k+1) +(k-1)\tanh(\{\sigma_{NN}T_A T_A-g\}/\delta)\right]$ with the total inelastic nucleon--nucleon cross-section $\sigma_{NN}$, and $g = \delta = 0.5$ fm$^{-2}$. This smoothly connects the FS/BJ-evolved pQCD energy density $\varepsilon(p_{\rm sat}^{\rm min})=C (T_A T_A)^k$ to the binary profile $\varepsilon \propto T_AT_A$ at the dilute edge. 

For the fluid-dynamical evolution, we use the state-of-the art 2+1 D setup previously employed in
Ref.~\cite{Niemi:2011ix, Niemi:2012ry, Niemi:2012aj}, assuming longitudinal boost invariance, a zero net-baryon density and thermalization at $\tau_0$. The equations of motion are given by the conservation laws for energy and momentum,  $\partial_{\mu}T^{\mu \nu }=0$. The evolution equation of the shear-stress tensor $\pi^{\mu\nu}=T^{\langle\mu\nu\rangle}$ is given by transient relativistic fluid dynamics~\cite{IS,Denicol:2012cn,Molnar:2013lta}, 
\begin{equation}
\tau_\pi \dot{\pi}^{\left<\mu\nu\right>} + \pi^{\mu\nu} =
2\eta \sigma^{\mu \nu}
- c_1 \pi^{\mu\nu} \theta
- \left(c_2 \sigma^{\left< \mu \right.}_{\,\,\,\, \lambda} - c_3 \pi^{\left< \mu \right.}_{\,\,\,\, \lambda}\right) \pi^{\left.\nu \right> \lambda} \nn
\end{equation}
where the co-moving time derivative $u^{\mu }\partial _{\mu }$ is denoted by the dot, $\eta $ is the shear viscosity coefficient, $\sigma^{\mu \nu}=\partial^{ \left<\mu \right.}u^{\left. \nu\right>}$ is the shear tensor, $\theta = \partial_\mu u^\mu$ is the expansion rate, and the angular brackets $\left<\right>$ denote the symmetrized and traceless projection, orthogonal to the fluid four-velocity $u^\mu$. The coefficients of the non-linear terms are taken to be 
$c_1 = 4\tau_\pi/3$, $c_2 = 10\tau_\pi/7$  and $c_3 = 9/(70 p)$, where $p$ is the thermodynamic pressure and $\tau_\pi=5\eta/(\varepsilon+p)$. For details of the numerical algorithm, see Refs.~\cite{Niemi:2012ry, Molnar:2009tx}.

The hadron spectra are calculated with the Cooper-Frye freeze-out procedure 
\cite{Cooper:1974mv} by using Israel's and Stewart's 14-moment ansatz
for the dissipative correction to the local equilibrium distribution
function, $\delta f_i = f_{0i} p_i^\mu p_i^\nu \pi_{\mu\nu}/[2T^2\left(\varepsilon+p\right)]$,
where $f_{0i} = \left\{\exp\left[\left(u_\mu p_i^\mu-\mu_i\right)/T\right]
\pm 1 \right\}^{-1}$, with the index $i$ indicating different hadron species and $p_i^\mu$ the
4-momentum of the corresponding hadron. The freeze-out temperature
is here always $T_{\rm dec} = 100$ MeV. After calculating the thermal spectra, we include the
contribution from all 2- and 3-particle decays of unstable
resonances in the EoS.

We use the lattice QCD and hadron resonance gas (HRG) based EoS 
$s95p$-PCE-v1~\cite{Huovinen:2009yb} with a chemical freeze-out temperature 
$T_{\rm chem} = 175$~MeV. Although the rather high $T_{\rm chem}$ 
leads to an overabundance of protons, it however reproduces the low-$\ptr$ region
of the $\ptr$-spectra much better than e.g.\ $T_{\rm chem} = 150$~MeV. 

\begin{figure}[!h]
\epsfysize 5.4cm \epsfbox{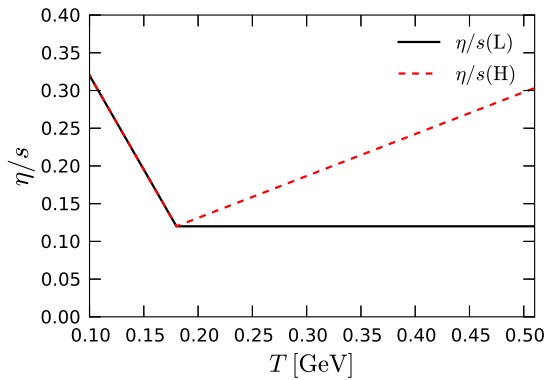} 
\vspace{-0.5cm}
\caption{\protect (Color online) Shear viscosity-to-entropy ratio as a function of temperature.}
\label{fig:etapers}
\end{figure}

For a rough but realistic (non-constant \cite{Csernai:2006zz}) shear viscosity description, we assume the ratio $\eta/s$ to decrease linearly as a function of temperature in the hadronic phase, be in a minimum at the matching-temperature 180 MeV of the HRG/QGP phases in the used EoS, and either to increase or stay constant vs. $T$ in the QGP phase \cite{Niemi:2011ix, Niemi:2012ry}. Fig~\ref{fig:etapers} shows the $\eta/s(T)$ which in our framework best reproduce the  $v_2$ coefficients simultaneously at RHIC and LHC.

At this point, we have a fixed framework with four \textit{correlated} unknowns, $\{K_{\rm sat}, \beta, {\rm BJ/FS}, \eta/s(T) \}$, to be determined using the LHC and RHIC data on the centrality dependence of the charged particle multiplicities, $p_T$ spectra and $v_2$. 
We proceed by scanning the  parameters  $K_{\rm sat}={\cal O}(1)$, $\beta \in [0,1]$ and  $\eta/s(T)$. In particular, we vary the minimum value and slopes of $\eta/s(T)$, keeping its general shape as in  Fig.~\ref{fig:etapers}. Both the BJ and FS prethermal evolutions are considered.  In practice, for each fixed $\{\beta, {\rm BJ/FS}, \eta/s(T) \}$, the remaining parameter $K_{\rm sat}$ is always tuned such that the multiplicity in the $0-5$ \% most central collisions at the LHC is reproduced. 

\begin{figure*}[!]
\epsfxsize 8.9cm \epsfbox{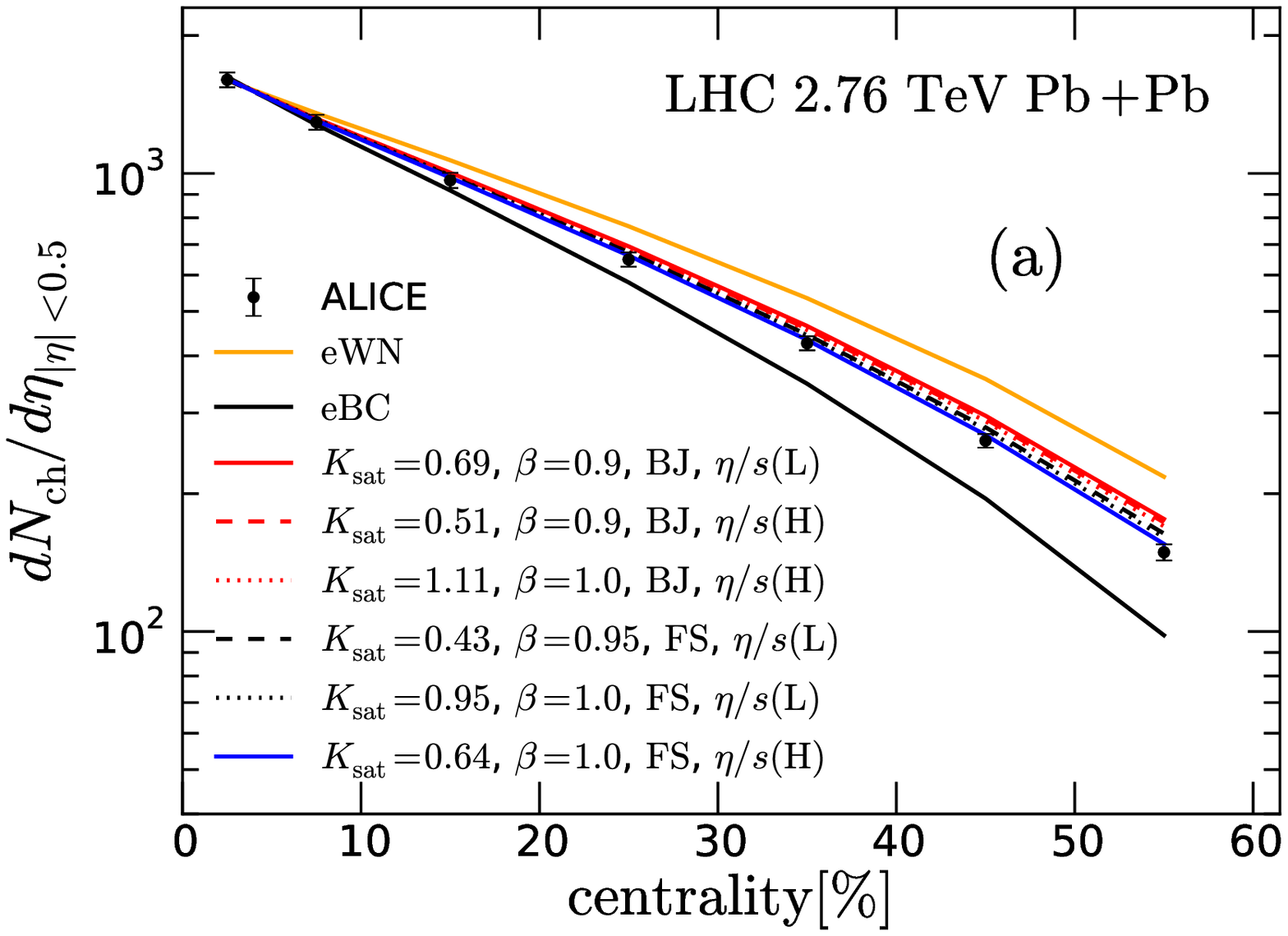} 
\epsfxsize 8.9cm \epsfbox{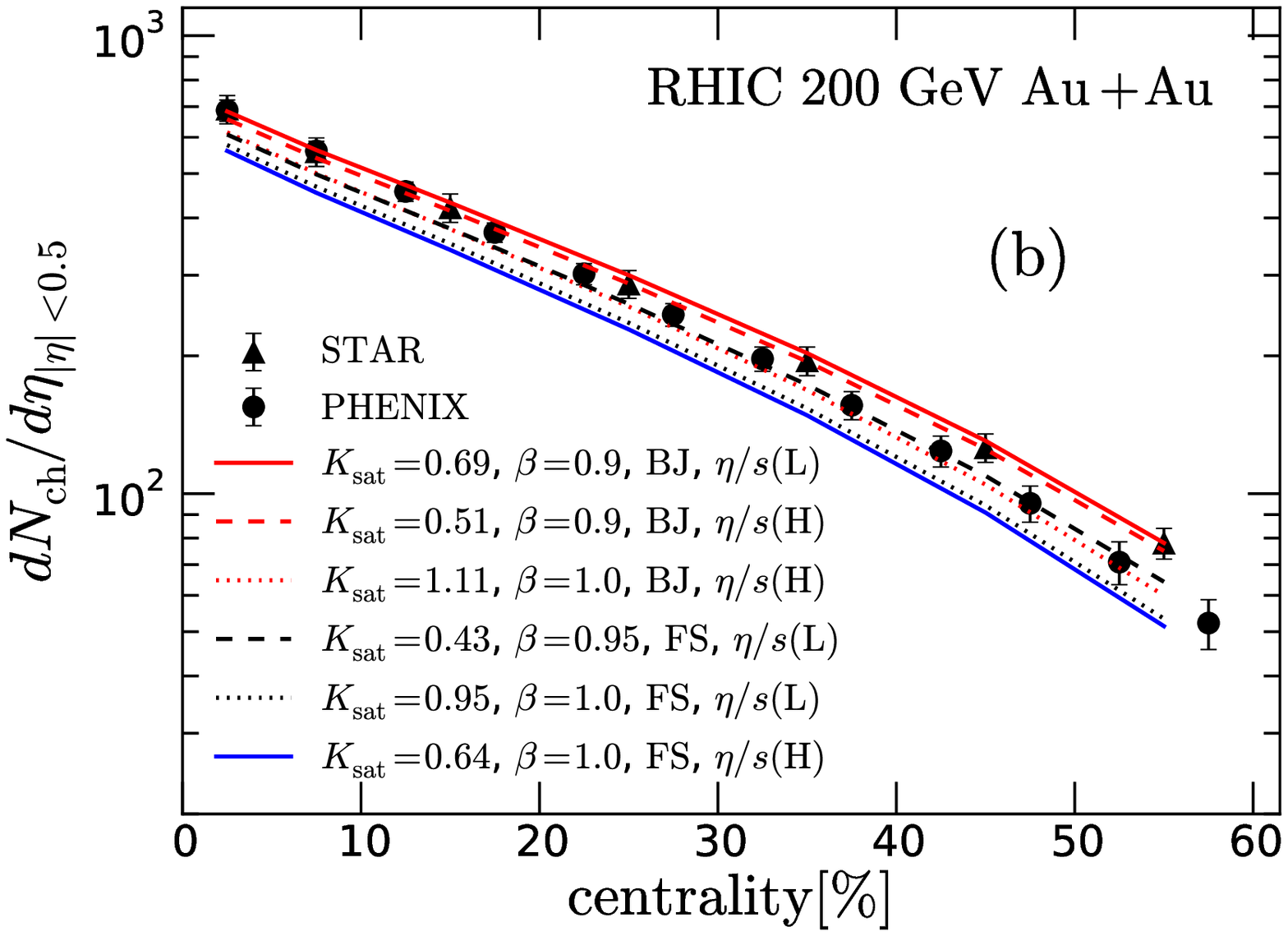} 
\epsfxsize 8.9cm \epsfbox{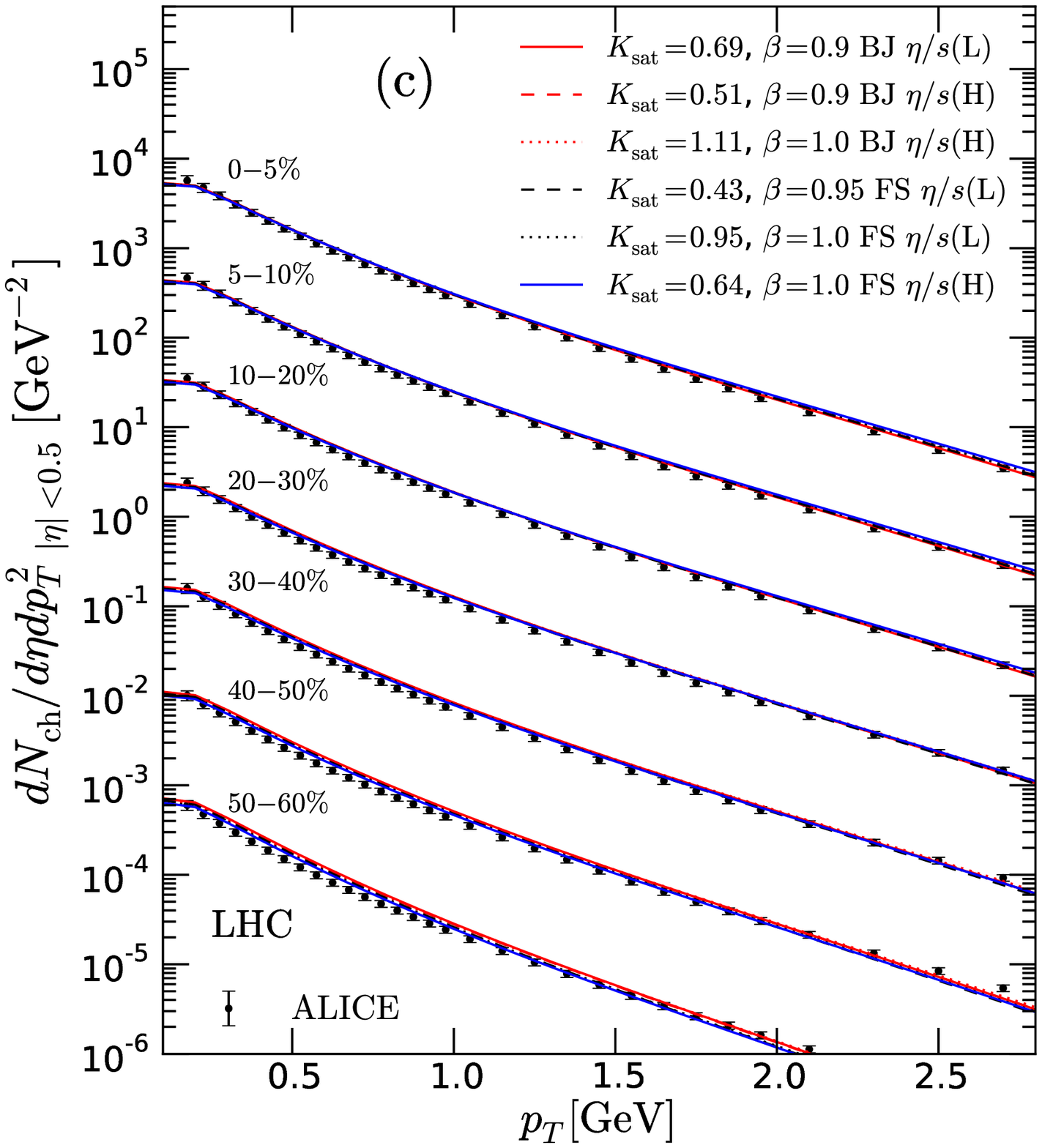} 
\epsfxsize 8.9cm \epsfbox{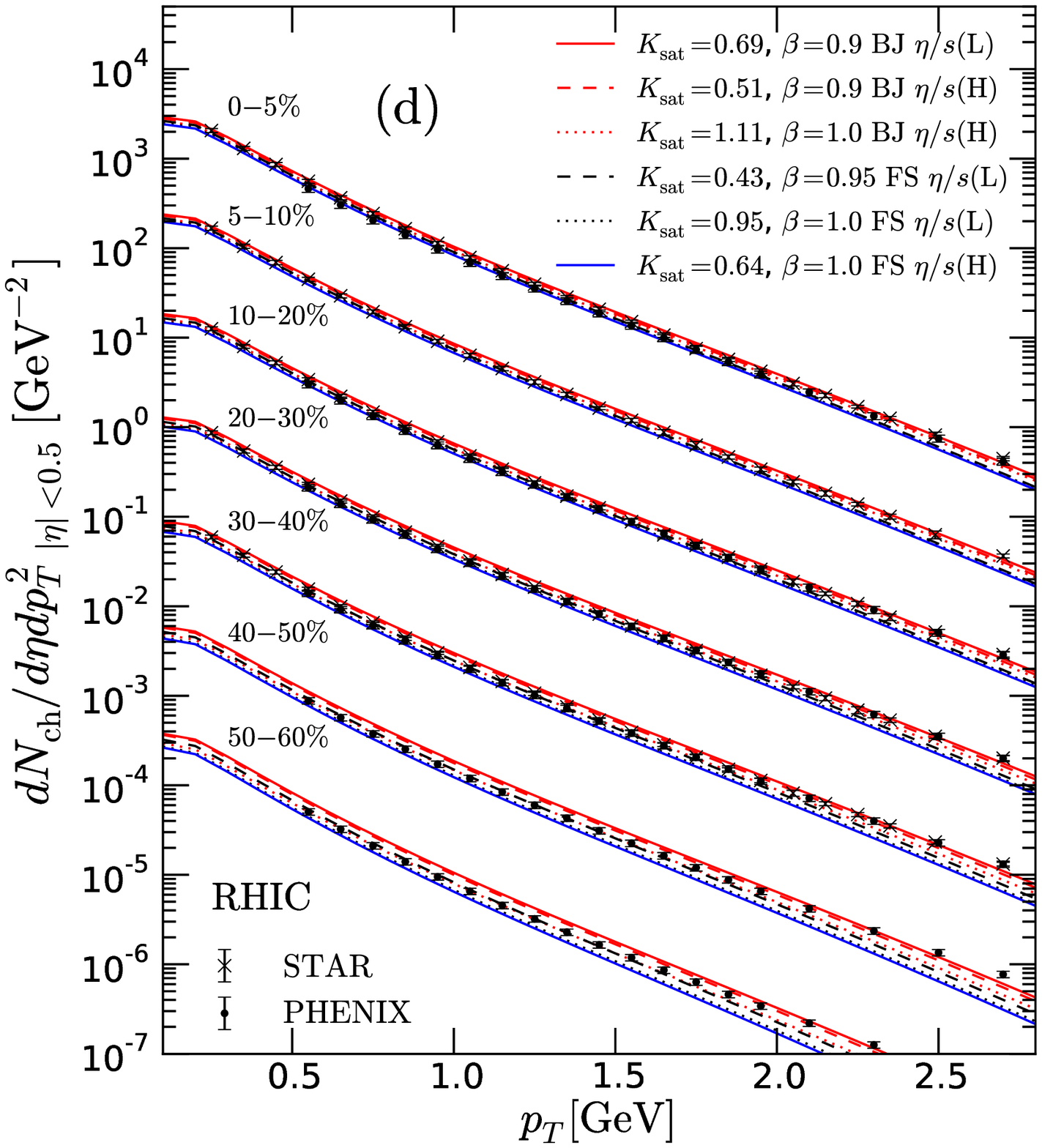} 
\epsfxsize 8.9cm \epsfbox{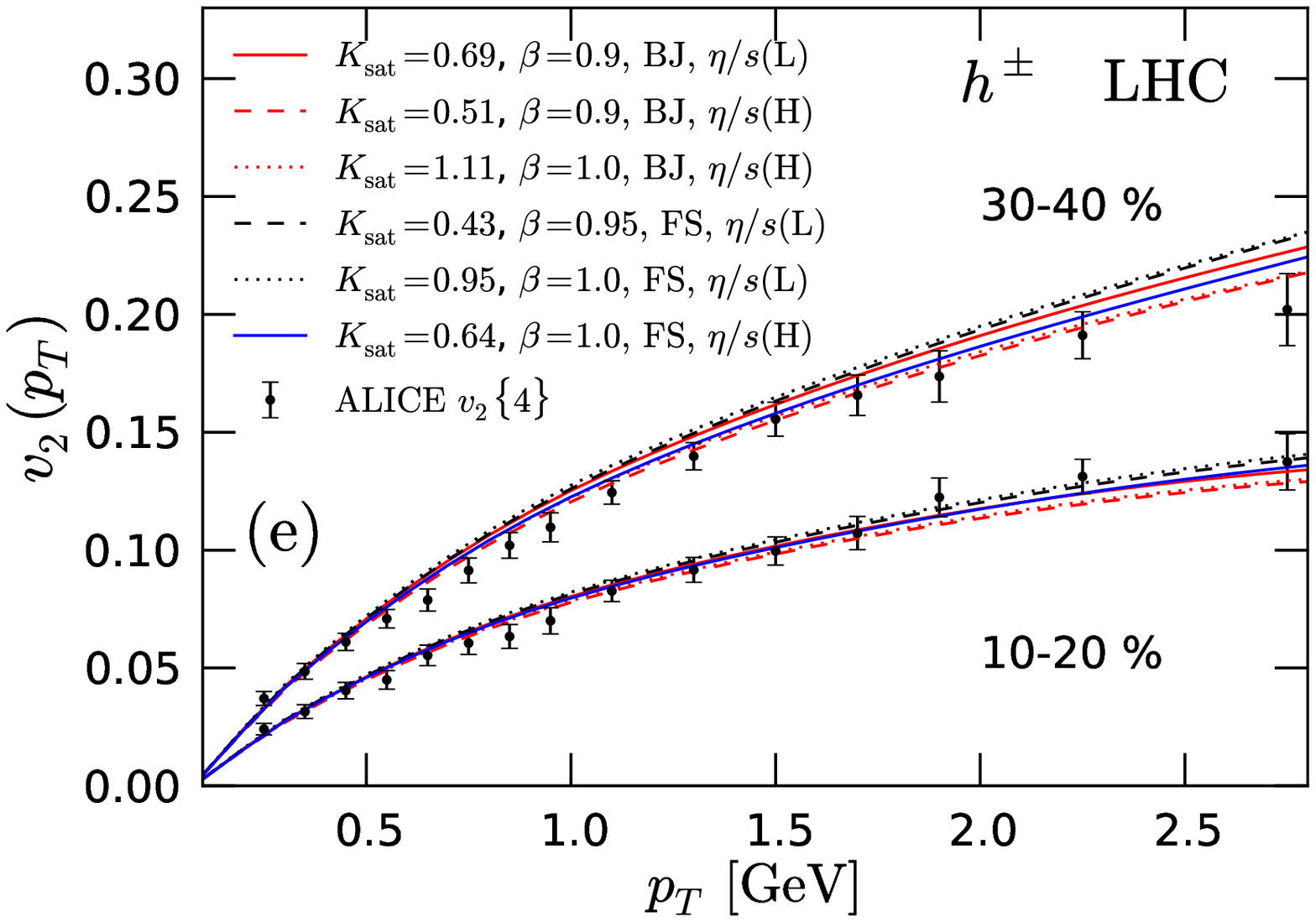} 
\epsfxsize 8.9cm \epsfbox{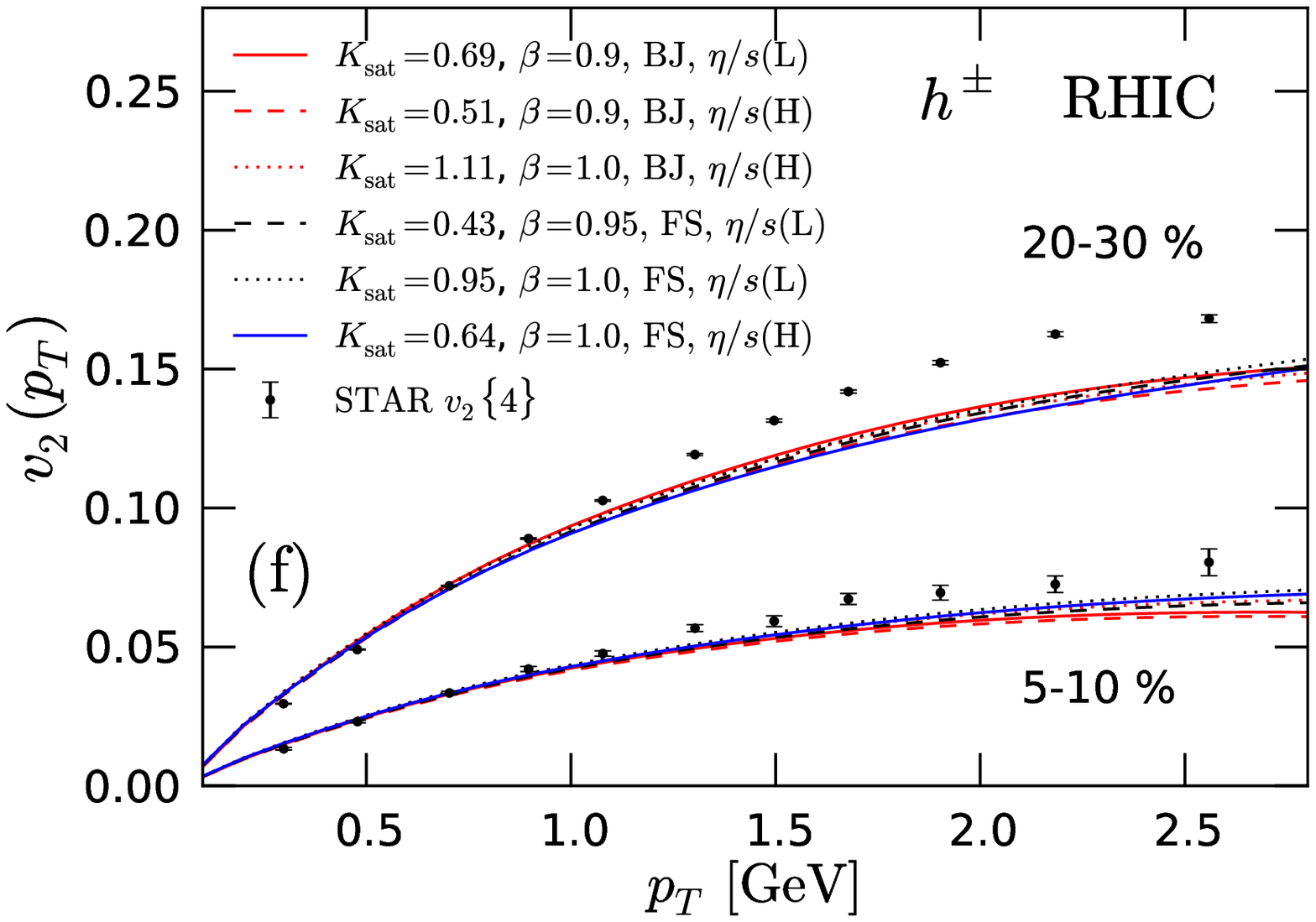} 
\caption{\protect (Color online)
Centrality dependence of the charged hadron multiplicity at the LHC (a) and RHIC (b). 
Transverse momentum spectra of charged hadrons at the LHC (c) and RHIC (d), in the same centrality classes as the ALICE data in panel (a), and scaled down by increasing powers of 10.
Elliptic flow coefficients $v_2(p_T)$ at the LHC (e) and RHIC (f), compared with the measured 4-particle cumulant $v_2\{4\}(p_T)$. Labeling of the theory curves in each panel is identical, and the parameter sets $\{K_{\rm sat}, \beta, {\rm BJ/FS}, \eta/s(T) \}$ are indicated. The labels H and L refer to Fig.~\ref{fig:etapers}.
}
\label{fig:results}
\end{figure*}

In Fig.~\ref{fig:results}a we show the computed centrality dependence of the  charged hadron multiplicity in Pb+Pb collisions at $\sqrt{s_{NN}} = 2.76$ TeV  compared with the ALICE data~\cite{Aamodt:2010pb}. As demonstrated here, several sets $\{K_{\rm sat}, \beta, {\rm BJ/FS}, \eta/s(T) \}$ give a good agreement with the measurement. However, the data clearly favours $\beta \sim 1$ and slightly the FS scenario over the BJ.  For comparison, we also show the results obtained with the usual (non-saturation) eBC and eWN Glauber model initial states~\cite{Kolb:2001qz}.

In Fig.~\ref{fig:results}b we show the multiplicities for Au+Au collisions at $\sqrt{s_{NN}} = 200$ GeV, using the same parameter sets $\{K_{\rm sat}, \beta, {\rm BJ/FS}, \eta/s(T) \}$ as in panel \ref{fig:results}a,  and compare with the PHENIX \cite{Adler:2004zn} and STAR \cite{Abelev:2008ab} data.  We note that although the RHIC data would seem to favour a slightly smaller $\beta$ and the BJ case, the overall simultaneous agreement at RHIC and LHC is rather good. 

As long as the centrality dependence of the multiplicity is described,  all the scenarios studied here give a very good description of the charged hadron $\ptr$-spectra. More relevant parameters in this case are $T_{\rm chem}$ and $T_{\rm dec}$ which here are kept unchanged from RHIC to LHC.
The obtained $p_T$ spectra are shown in Fig.~\ref{fig:results}c for the LHC and in Fig.~\ref{fig:results}d for RHIC. The data  are from Refs.~\cite{Abelev:2012hxa} and Ref.~\cite{Adams:2003kv, Adler:2003au}, correspondingly. 

In Figs.~\ref{fig:results}e and \ref{fig:results}f we show the elliptic  flow coefficients $v_2(\ptr)$ at the LHC and RHIC, respectively. The data are from ALICE \cite{Aamodt:2010pa} and STAR \cite{Bai}. The $v_2(\ptr)$ coefficients depend strongly on the  $\eta/s$ parametrization, and \textit{e.g.} an ideal fluid description (not shown) does not give a correct $v_2(\ptr)$. By scanning the $\eta/s(T)$ as explained above, while keeping $K_{\rm sat}$ of order 1,  we observed that a good simultaneous agreement with the measurements is obtained with the cases shown in  Fig.~\ref{fig:etapers}. We emphasize that at RHIC, where the flow gradients are larger,  one should require the agreement in particular in the small-$p_T$ region, where  the dissipative corrections to the particle distributions do not grow unphysically large. Note especially that since $\eta/s(T)$ is considered as a material property, it must not be changed between different collision systems. 

To conclude, we computed the energy density profiles and formation times of the produced QGP at the LHC and RHIC in a new NLO-improved pQCD + local saturation framework of considerable predictive power. The subsequent evolution of these initial conditions was described with dissipative fluid dynamics. Identifying the framework uncertainties, a good global agreement with the measured centrality dependence of the low-$p_T$ bulk observables was obtained simultaneously at the LHC and RHIC. In particular, we were able to constrain the $\eta/s(T)$ parametrization simultaneously by all these data. In the future, we will extend this analysis to include event-by-event fluctuations.

\textit{Acknowledgements.}
This work was financially supported by the Wihuri foundation (RP) and the Academy of Finland, projects 133005 (KJE) and 267842 (KT). We thank I.~Helenius, T.~Lappi, H.~M\"antysaari and H.~Paukkunen for useful discussions, and CSC-IT Center Science for supercomputing time.


\begin{thebibliography}{50} 

\bibitem{Heinz:2013th} For a recent review and references, see
  U.~W.~Heinz and R.~Snellings,
  Annu.\  Rev.\  Nucl.\  Part.\  Sci.\  {\bf 63}, 123 (2013).

\bibitem{Romatschke:2007mq} 
  P.~Romatschke and U.~Romatschke,
  Phys.\ Rev.\ Lett.\  {\bf 99}, 172301 (2007).

\bibitem{Luzum:2008cw} 
  M.~Luzum and P.~Romatschke,
  Phys.\ Rev.\ C {\bf 78}, 034915 (2008)
  [Erratum-ibid.\ C {\bf 79}, 039903 (2009)].
  
\bibitem{Schenke:2010rr} 
  B.~Schenke, S.~Jeon and C.~Gale,
  Phys.\ Rev.\ Lett.\  {\bf 106}, 042301 (2011);
  Phys.\ Rev.\ C {\bf 85}, 024901 (2012);
  Phys.\ Lett.\ B {\bf 702}, 59 (2011).

\bibitem{Gale:2012rq} 
  C.~Gale  {\it et al.},
  Phys.\ Rev.\ Lett.\  {\bf 110}, 012302 (2013).
  
\bibitem{Song:2010mg} 
  H.~Song  {\it et al.},
  Phys.\ Rev.\ Lett.\  {\bf 106}, 192301 (2011)
  [Erratum-ibid.\  {\bf 109}, 139904 (2012)];
  Phys.\ Rev.\ C {\bf 83}, 054910 (2011)
  [Erratum-ibid.\ C {\bf 86}, 059903 (2012)].

\bibitem{Song:2011qa} 
  H.~Song, S.~A.~Bass and U.~Heinz,
  Phys.\ Rev.\ C {\bf 83}, 054912 (2011)
  [Erratum-ibid.\ C {\bf 87}, 019902 (2013)].

\bibitem{Shen:2010uy} 
  C.~Shen, U.~Heinz, P.~Huovinen and H.~Song,
  Phys.\ Rev.\ C {\bf 82}, 054904 (2010);
  Phys.\ Rev.\ C {\bf 84}, 044903 (2011).

\bibitem{Bozek:2009dw} 
  P.~Bozek,
  Phys.\ Rev.\ C {\bf 81}, 034909 (2010);
  Phys.\ Rev.\ C {\bf 85}, 034901 (2012).


\bibitem{Bozek:2012qs} 
  P.~Bozek and I.~Wyskiel-Piekarska,
  Phys.\ Rev.\ C {\bf 85}, 064915 (2012).

\bibitem{Niemi:2012ry} 
  H.~Niemi  {\it et al.},
  Phys.\ Rev.\ C {\bf 86}, 014909 (2012).
  
\bibitem{Niemi:2011ix} 
  H.~Niemi  {\it et al.},
  Phys.\ Rev.\ Lett.\  {\bf 106}, 212302 (2011).
    
\bibitem{Gribov:1984tu}
  L.~V.~Gribov, E.~M.~Levin and M.~G.~Ryskin,
  Phys.\ Rept.\  {\bf 100} (1983) 1.
  
\bibitem{Mueller:1985wy} 
  A.~H.~Mueller and J.~Qiu,
  Nucl.\ Phys.\ B {\bf 268}, 427 (1986).

\bibitem{McLerran:1993ni} 
  L.~D.~McLerran and R.~Venugopalan,
  Phys.\ Rev.\ D {\bf 49}, 2233 (1994).

\bibitem{Eskola:1996ce} 
  K.~J.~Eskola and K.~Kajantie,
  Z.\ Phys.\ C {\bf 75}, 515 (1997).
  
\bibitem{Eskola:1999fc} 
  K.~J.~Eskola, K.~Kajantie, P.~V.~Ruuskanen and K.~Tuo\-mi\-nen,
  Nucl.\ Phys.\ B {\bf 570}, 379 (2000).

\bibitem{Paatelainen:2012at}
	R.~Paatelainen, K.~J.~Eskola, H.~Holopainen and K.~Tuominen,
  Phys.\ Rev.\ C {\bf 87}, 044904 (2013).
  
\bibitem{Eskola:2009uj} 
  K.~J.~Eskola, H.~Paukkunen and C.~A.~Salgado,
  JHEP {\bf 0904}, 065 (2009).

\bibitem{Gelis:2010nm} 
  F.~Gelis, E.~Iancu, J.~Jalilian-Marian and R.~Venugopalan,
  Ann.\ Rev.\ Nucl.\ Part.\ Sci.\  {\bf 60}, 463 (2010).

\bibitem{Eskola:2001bf} 
  K.~J.~Eskola, P.~V.~Ruuskanen, S.~S.~R\"as\"anen and K.~Tuo\-minen,
  Nucl.\ Phys.\ A {\bf 696}, 715 (2001).
  
\bibitem{Eskola:2002wx} 
  K.~J.~Eskola, H.~Niemi, P.~V.~Ruuskanen and S.~S.~R\"a\-s\"anen,
  Phys.\ Lett.\ B {\bf 566}, 187 (2003).
  
\bibitem{Eskola:2005ue}
	K.~J.~Eskola  {\it et al.},
	Phys.\ Rev.\  C {\bf 72}, 044904 (2005).
  
\bibitem{Renk:2011gj} 
  T.~Renk, H.~Holopainen, R.~Paatelainen and K.~J.~Eskola,
  Phys.\ Rev.\ C {\bf 84}, 014906 (2011).
  
\bibitem{Eskola:2000xq} 
  K.~J.~Eskola, K.~Kajantie and K.~Tuominen,
  Phys.\ Lett.\ B {\bf 497}, 39 (2001).

\bibitem{Abelev:2008ab} 
  B.~I.~Abelev {\it et al.}  [STAR Collaboration],
  Phys.\ Rev.\ C {\bf 79}, 034909 (2009).

\bibitem{Eskola:1988yh} 
  K.~J.~Eskola, K.~Kajantie and J.~Lindfors,
  Nucl.\ Phys.\ B {\bf 323}, 37 (1989).

\bibitem{tuominen:2000}
	K.~J.~Eskola and K.~Tuominen,
	Phys.\ Let.\  B {\bf 489}, 329 (2000);
	Phys.\ Rev.\  {\bf D63}, 114006 (2001).

\bibitem{Pumplin:2002vw} 
  J.~Pumplin  {\it et al.}, 
  JHEP {\bf 0207}, 012 (2002).
  
\bibitem{helenius:2012wd}
  I.~Helenius, K.~J.~Eskola, H.~Honkanen and C.~A.~Salgado,
	JHEP\ {\bf 1207}, 073 (2012).

\bibitem{Kunszt:1992tn} 
  Z.~Kunszt and D.~E.~Soper,
  Phys.\ Rev.\ D {\bf 46}, 192 (1992).

\bibitem{Eskola:2001rx} 
  K.~J.~Eskola, K.~Kajantie and K.~Tuominen,
  Nucl.\ Phys.\ A {\bf 700}, 509 (2002).

\bibitem{Niemi:2012aj} 
  H.~Niemi, G.~S.~Denicol, H.~Holopainen and P.~Huovinen,
  Phys.\ Rev.\ C {\bf 87}, 054901 (2013).

\bibitem{IS}
	W.~Israel and J.~M.~Stewart,
	Proc.\ R.\ Soc.\ A {\bf 365}, 43 (1979);
  Ann. Phys. (N.Y.)\  {\bf 118}, 341 (1979).

\bibitem{Denicol:2012cn} 
  G.~S.~Denicol, H.~Niemi, E.~Molnar and D.~H.~Rischke,
  Phys.\ Rev.\ D {\bf 85}, 114047 (2012).

\bibitem{Molnar:2013lta} 
  E.~Molnar, H.~Niemi, G.~S.~Denicol and D.~H.~Rischke,
  arXiv:1308.0785 [nucl-th].

\bibitem{Molnar:2009tx} 
  E.~Molnar, H.~Niemi and D.~H.~Rischke,
  Eur.\ Phys.\ J.\ C {\bf 65}, 615 (2010).

\bibitem{Cooper:1974mv}
	F.~Cooper, G.~Frye,
	Phys.\ Rev.\  D {\bf 10}, 186 (1974).

\bibitem{Huovinen:2009yb}
  P.~Huovinen, P.~Petreczky,
  Nucl.\ Phys.\  {\bf A837}, 26 (2010).

\bibitem{Csernai:2006zz}
  L.~P.~Csernai, J.~I.~Kapusta, L.~D.~McLerran,
  Phys.\ Rev.\ Lett.\  {\bf 97}, 152303 (2006).

\bibitem{Aamodt:2010pb} 
  K.~Aamodt {\it et al.}  [ALICE Collaboration],
  Phys.\ Rev.\ Lett.\  {\bf 105}, 252301 (2010).

\bibitem{Kolb:2001qz}
	P.~F.~Kolb {\it et al.}, 
	Nucl.\ Phys.\  A {\bf 696}, 197 (2001).

\bibitem{Adler:2004zn} 
  S.~S.~Adler {\it et al.}  [PHENIX Collaboration],
  Phys.\ Rev.\ C {\bf 71}, 034908 (2005)
  [Erratum-ibid.\ C {\bf 71}, 049901 (2005)].

\bibitem{Abelev:2012hxa} 
  B.~Abelev {\it et al.}  [ALICE Collaboration],
  Phys.\ Lett.\ B {\bf 720}, 52 (2013).

\bibitem{Adams:2003kv} 
  J.~Adams {\it et al.}  [STAR Collaboration],
  Phys.\ Rev.\ Lett.\  {\bf 91}, 172302 (2003).
    
\bibitem{Adler:2003au} 
  S.~S.~Adler {\it et al.}  [PHENIX Collaboration],
  Phys.\ Rev.\ C {\bf 69}, 034910 (2004).

\bibitem{Aamodt:2010pa} 
  K.~Aamodt {\it et al.}  [ALICE Collaboration],
  Phys.\ Rev.\ Lett.\  {\bf 105}, 252302 (2010).
  
\bibitem{Bai}
 	Y.~Bai, Ph.D. Thesis, Nikhef and Utrecht University, The Netherlands (2007);
  A.~Tang  [STAR Collaboration],
  arXiv:0808.2144 [nucl-ex].

\end{thebibliography}
\end{document}